# Empirical quantification of rockfall reach probability: objective determination of appropriate topographic descriptor

M. Peruzzetto[1], B. Colas[2], C. Levy[1], J. Rohmer[1], F. Bourrier[3]

[1] BRGM, Orléans, France
[2] BRGM, Montpellier, France
[3] Université Grenoble Alpes, INRAE, ETNA, 38000, Grenoble, France

SUMMARY: For rockfall hazard assessment on areas more than several km² in size, the quantification of runout probability is usually done empirically. Classical methods use statistical distributions of reach or energy angles derived from rockfall databases. However, other topographic descriptors can be derived from the topographic profiles along the rockfall path. Using a database of more than 4,000 profiles of rockfall paths, we determine which topographic descriptors are most appropriate for reach probability estimation, by comparing their statistical distributions for rockfall stopping points, and for points the rocks overtook. We show that the curvilinear length of the propagation path, and the area under the propagation path can improve propagation estimations, especially when they are computed only along the final portion of the propagation path. This is illustrated by comparing an experimental distribution of rockfall stopping points to the estimated distributions.

## Introduction

The estimation of rockfall runout probability is a key step of rockfall hazard assessment for delimiting areas of low, moderate and high hazard levels below rocky cliffs. The most sophisticated tools used to analyse rockfall propagation are 3D trajectography numerical models. However they rely sometimes on many user-defined parameters that can be difficult to calibrate. They also require significant computing resources that are often incompatible with the size of the study area (e.g. several km² for regulatory hazard maps) and IT resources at disposal. Thus, practitioners commonly rely on empirical methods, with scalar topographic descriptors derived from 2D topographic profiles of rock propagation paths. A widely used indicator is the ratio $\mu = \tan(\delta)$ between the drop height $H$ and the travel distance $L$, with $\delta$ the corresponding angle. $\delta$ is called the angle of reach when $L$ is the curvilinear length of the real path, and the angle of energy when $L$ is the horizontal distance between the starting and stopping points. By using empirical distributions of $\mu$ derived from rockfall databases, probabilistic runout maps can be easily derived (e.g. Jaboyedoff et al., 2003). It was also suggested that combining the energy angle with the area under the propagation path (normalized by the drop height) could also provide good results (MEZAP, 2021). This method is currently implemented by the BRGM to derive regulatory rockfall hazard maps (Levy et al., 2021).

The objective quantification of the precision of an empirical method for runout probabilistic prediction is difficult. How can we find a topographic descriptor that is best representative of rockfall stopping points, and not of the general shape of slopes the rocks propagate on? To the knowledge of the authors, this problem has never, or at least only scarcely, been addressed in the literature. We suggest that an appropriate criterion is the comparison between the statistical distributions of topographic descriptors computed for actual rockfall stopping points, and the distributions for other points, as explained in the next section.

**Data and methods**

We use a database of 4,015 topographic profiles $z = z(x)$ connecting with straight lines the initiation and stopping points of rockfalls, derived from 15 different catchments in Europe. For every point with coordinate $x$ along a profile $z = z(x)$, we compute the distance $L(x)$ to the initiation point, the drop height $H(x) = z(0) - z(x)$, $A(x)$ the area under the profile and $C(x)$ the curvilinear length of the profile (both computed between 0 and $x$ coordinates, on the vertically translated profile $z_x(r) = z(r) - z(x)$), and non-dimensional indicators given in Table 1.

Table 1. Non-dimensionnal topographic indicators

| Topographic descriptor | $\mu$ | $A_n$ | $A_n^L$ | $A_n^H$ | $C_n$ | $C_n^L$ | $C_n^H$ |
|---|---|---|---|---|---|---|---|
| Definition | $H/L$ | $A/(HL)$ | $A/L^2$ | $A/H^2$ | $C/\sqrt{L^2+H^2}$ | $C/L$ | $C/H$ |

The indicators are also computed by considering only the last $d$ meters before a given point. For instance, $\mu(x) = \frac{z(0)-z(x)}{x}$ and $\mu_{(d)}(x) = \frac{z(x-d)-z(x)}{d}$. As a result, for each indicator, we derive a sample of 4,015 points corresponding to rockfall stopping points (rockfall database), and a sample of almost $10^6$ points corresponding to all $(x, z(x))$ couples from the 4,015 profiles that rockfalls have overtook (profiles database). The similarity between distributions is quantified by computing the Wasserstein (or Kantorovitch) distance $D_W$ between bootstrapped samples of the normalized rockfalls and profiles distributions (taking the profiles distribution as a reference). The Wasserstein distance corresponds to the optimal value of the optimal transport problem: it is all the more important as distributions are different. Thus, topographic descriptors that allow to better discriminate between rockfall stopping points and points that they overtake are associated to higher values of $W_d$.

Then, to estimate reach probabilities on a new profile $z = z(x)$ with a given topographic indicator $y$, we first fit a non-parametric normal kernel density function $f$ to the distribution of $y$ for rockfall stopping points. Assuming that the rockfall is initiated at $x = 0$ and stops before $x = x_m$, the probability that the rockfall has a travel distance $X$ larger than $x$ is estimated as:

$$P(X > x) = 1 - \frac{1}{C}\int_0^x f(y(r))dr \text{ with } C = \int_0^{x_m} f(y(r))dr \quad (1)$$

We compare estimated reach probabilities to observations from two experiments carried out in Dole (Jura, France) where 47 and 56 blocks with mass varying from 100 to 1,300 kg were released from two locations. This empriccaly derived probability distribution is termed as observed distribution in the following.

**Results**

As shown in Figure 1a, the topographic indicator that allows to best discriminate between rockfall stopping points and other points, when the full profile is considered, is $A_n^L$ ($D_W = 0.36$). The difference between distributions is even more pronounced for $A_{n\,(30)}^H$ and $C_{n\,(30)}^H$ ($D_W = 1.5$ and $D_W = 4.3$ respectively).

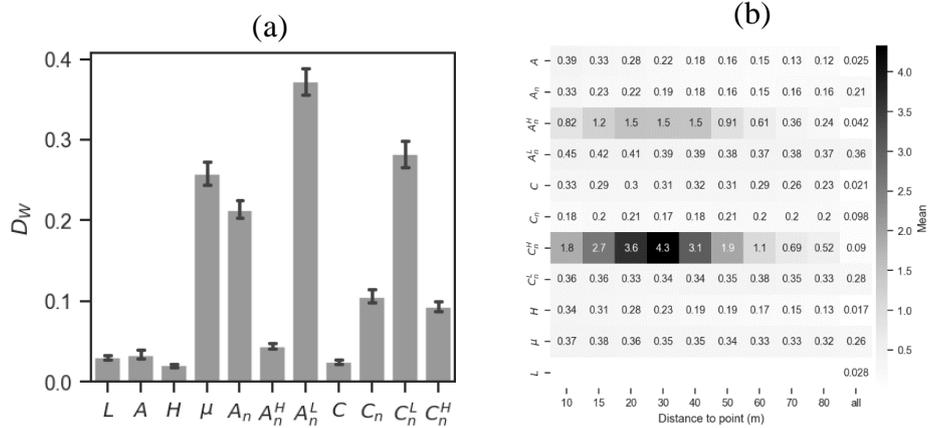

Figure 1 : Dissimilarity between distributions for rockfall stopping points, and other points, for different topographic indicators, measured with the mean of the bootstrapped Wasserstein distance $D_W$. Topographic indicators are computed using the whole profile in (a) and in the last column of (b), and using only the last 10 to 80 meters in (b).

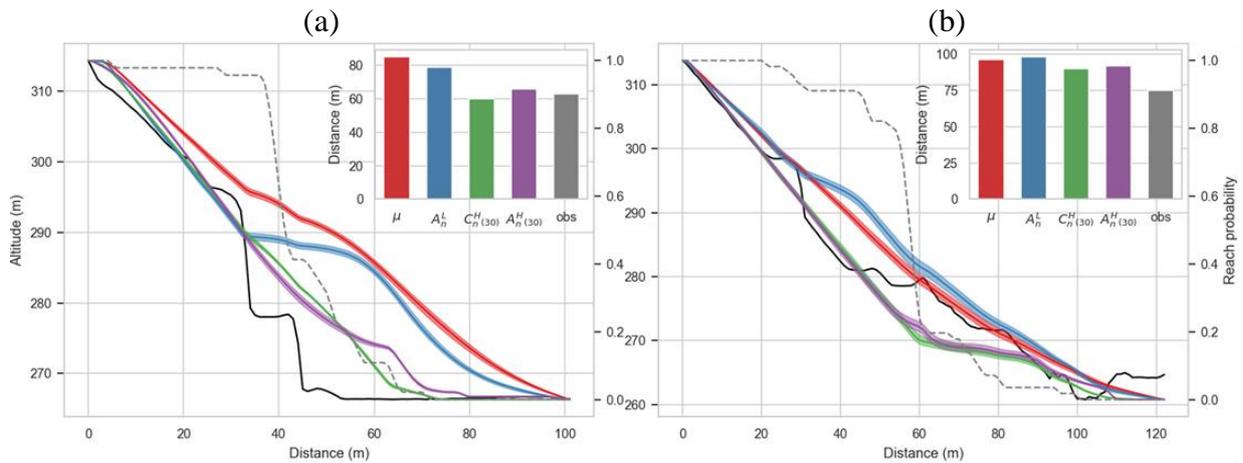

Figure 2 : Comparison between observed (grey dashed line) and estimated (coloured lines with 95% confidence intervals) travel distance exceedance probabilities, for two profiles (a and b) in Dole (Jura, France). The topographic profile is given in black. The insert gives the distance corresponding to an exceedance probability of 0.1. $C^H_{n(30)}$ and $A^H_{n(30)}$ (green and purple) are computed on the last 30 meters.

In Figure 2 a practical application is presented on two topographic profiles with several slope breaks, and an observed distribution of rockfall stopping points. The observed reach probability decreases sharply in response to slope breaks (see e.g. in Figure 2a between *x*=30m and *x*=40m). In comparison, the reach probability derived from $\mu$ and $A^L_n$ decreases relatively regularly along the profile: it is under-estimated for the first 40 to 60 m, and then over-estimated. As a result, the travel distance associated to the 10% reach probability is over-estimated by 20 to 25 m (see inserts in Figure 2). On the contrary, the estimations deduced from $C^H_{n\,(30)}$ and $A^H_{n\,(30)}$ are in better agreement with observations (within 5 meters in Figure 2a). For these indicators, the reach probability is also under-estimated in the first 60 meters, but is then in good agreement with observations.

**Discussion**

Our results seem to validate the proposed methodology for determining topographic indicators that are characteristic of rockfall stopping points. However, the tested profiles are not representative of all rockfall propagation conditions. In particular, the drop height in the application in 50 m at maximum, which is not representative of high mountain cliffs. Testing the methodology on such a cliff with a debris fan at its foot, and an associated database of observed rockfalls, would be interesting. In the meantime, other possible improvements include the analysis of the combined distribution of multiple topographic indicators. For instance, using the combined distribution of $C^H_{n\,(30)}$ and $L$ allows a better match between the observed and estimated reach probability in the first 60 meters of the Dole profiles, but has limited influence on the results at further distances. Using the combined distribution of $C^H_{n\,(30)}$ and $A^H_{n\,(30)}$ does not significantly improve the results.

In the perspective of deriving the most realistic reach probabilities on a given profiles, some difficulties are yet to be overcome. First, it should be noted that the proposed estimation of the reach probability (which is classical in the literature) is only based on the probability to have a certain value of a topographic index, given a range of possible value. But this range of values does not define unequivocally the profile itself, neither does a topographic index corresponds necessarily to a single point along the profile. Besides, a descriptor may work well on average for various types of profiles (as indicated by high values of $D_w$), but another indicator could be more appropriate for a specific profile. This problem can be partly overcome by filtering the distribution used to compute the reach probability. For instance, using the distribution of $C^H_{n\,(30)}$ for rockfall stopping points with travel distance below 150 m allows to improve the estimation of the travel distance with 10% reach probability for the test profile in Figure 2b.

A significant methodological result of our work is that topographic indicators computed only on a portion of the profile are more effective to derive reach probabilities that indicators computed on the whole profile. This highlights the role of the local topography in rockfall propagation, and especially slope breaks. In practice, using partial topographic descriptors (such as $A^H_{n\,(30)}$) allows to estimate more realistic low reach probabilities than descriptors integrating the whole path.

**Conclusion**

This work contributes to improving rockfall propagation hazard assessment with empirical methods, in particular for extensive studies on large surfaces. By computing various topographic descriptors along profiles from a database of rockfall propagation paths, we show that the curvilinear profile length on the last 30 meters, normalized by the drop height, is a satisfying descriptor to estimate travel distances. It allows to improve low reach probability estimations in comparison to indicators integrating the whole profile, such as the energy angle. Future developments could explore four main directions : (i) using the combined distributions of multiple topographic descriptors from a subset of topographic profiles to improve reach probability description on a given profile, (ii) extending the current profile database and testing the methodology on new sites, (iii) conducting a formal analysis of reach probability estimation to estimate and reduce the bias associated to current estimation methods, and (iv) analyzing the influence of block volume on travel distance.

**References**

Jaboyedoff, M., & Labiouse, V.. 2003. 'Preliminary Assessment of Rockfall Hazard Based on GIS Data'. In 10th International Congresson Rock Mechanics ISRM 2003 – Technology Roadmap for Rockmechanics, 575–78. Johannesburgh, South Africa.

Lévy, C., Colas, B., Rohmer, J., & Berger, F. (2021). ELANA (Energy Line Angle Normalized Area): un outil d'aide à la cartographie de la propagation des chutes de blocs basée sur la méthode de la ligne d'énergie à différentes échelles. In 5th RSS Rock Slope Stability Symposium.

MEZAP. 2021. 'Guide Technique MEZAP'. Caractérisation de l'aléa rocheux dans le cadre d'un Plan de Prévention des Risques Naturels (PPRn) ou d'un Porter a connaissance (PAC). Collection Scientifique et Technique. Orléans: BRGM.